\documentstyle[prl,aps,preprint,tighten,floats]{revtex}

\newcommand{\be}{\begin{equation}}
\newcommand{\ee}{\end{equation}}
\newcommand{\bea}{\begin{eqnarray}}
\newcommand{\eea}{\end{eqnarray}}
\newcommand{\ben}{\begin{enumerate}}
\newcommand{\een}{\end{enumerate}}
\newcommand{\bit}{\begin{itemize}}
\newcommand{\eit}{\end{itemize}}
\newcommand{\la}[1]{\label{#1}}
\newcommand{\fract}[2]{{\textstyle\frac{#1}{#2}}}

\newcommand{\eq}[1]{Eq.~(\ref{#1})}

\newcommand{\eqs}[2]{eqs.~(\ref{#1}) and (\ref{#2})}

\newcommand{\half}{\fract{1}{2}}

\newcommand{\Tr}{\,\mbox{Tr}\,}

\newcommand\mnl{{\mu\nu\lambda}}

\newcommand{\rd}[1]{\mathop{{\rm d}#1}}

\input BoxedEPS
\SetRokickiEPSFSpecial  
\HideDisplacementBoxes

\begin{document}
\draft

\preprint{\vbox{MIT-CTP-3077
                                        \null\hfill\rm February 2001}}

\title{The Spin Structure of the Nucleon:\\
Theoretical Overview\\[1ex]
{\small\it  Presented at the\\
Second Workshop on Physics with a Polarized-Electron Light-Ion Collider 
(EPIC)\\
September 14--16, 2000 at 
MIT, 
Cambridge, Massachusetts, USA}
}

\author{R.L.~Jaffe}

\address{Center for Theoretical Physics and Department of Physics \\
Laboratory for Nuclear Physics
Massachusetts Institute of Technology\\
Cambridge, Massachusetts 02139}

\maketitle

\begin{abstract}\noindent
I review what is known about the quark and gluon spin distributions in the
nucleon. I discuss in some detail (a)~the existence of sum rules for angular
momentum; (b)~the interpretation and possible measurement of the nucleon's
transversity distributions; and (c)~the uses of spin-dependent fragmentation
functions.
\end{abstract}

\section{Introduction}

The modern era in QCD spin physics dates from the 1987 discovery by
the European Muon Collaboration that only about 30\% of the proton's
spin is found on the spin of quarks~\cite{Ashman:1988hv}.  Since then, particle and
nuclear physicists have dreamt of facilities where QCD spin physics
could be explored in detail.  The recent commissioning of polarized
$pp$ physics at RHIC is the first of these to be realized~\cite{Spin2000-Saito}.
Our topic -- a polarized $ep$ collider in the energy regime where
perturbative QCD meets confinement -- is a necessary complement to
$\overrightarrow{\rm RHIC}$ and the natural next step in unravelling
the mysteries of quark confinement in QCD.

Among friends, I do not need to belabor the case for studying QCD
at the boundary between the confining and perturbative domains.  Two
brief comments will suffice: First, quantum chromodynamics is the only
nontrivial quantum field theory that we are certain describes the
real world; and second, we need further experimental input to
understand the highly complex QCD bound states that compose
matter.

This workshop focuses on spin.  While spin is an important degree of
freedom, it is not the only important probe of confinement in deep
inelastic processes.  Flavor, twist, and quark mass dependence 
(through the substitution $u\to d\to s\to c\to b\to t$) yield different
and complementary insights into the structure of QCD bound states. 
Spin, however, is today's topic.

Of course it is impossible to cover the breadth of this field in a
single talk.  Fortunately, others will address important subjects
in detail later at this meeting.  Instead, I will make a brief
survey of the present situation, emphasizing our present understanding 
of quark and gluon distribution functions, and then focus on three issues of
current interest:
\begin{itemize}
	\item Is there an ``angular momentum sum rule'' and is it 
experimentally testable?
	\item What is transversity and why is it interesting?
	\item Why are fragmentation functions interesting and useful in the 
	study of spin in QCD?
\end{itemize}

\section{The Present Situation}

Polarization effects in QCD present a complex picture.  Asymmetries
need to be explained, but sometimes even if we cannot understand them,
we can use them to probe other issues or isolate other important
effects.  Many striking asymmetries occur in the low energy or nuclear
domain where we have few theoretical insights into
QCD~\cite{Krisch:1998nw}. Most recent progress has occurred where the
deep inelastic and soft domains overlap -- the world of parton
distribution and fragmentation functions.  Here, spin effects help
elucidate the puzzling nature of hadrons and here is where I will
concentrate.

\subsection{Recent Events}

To set the stage for the workshop, here are lists of recent
developments in experiment and theory, and a menu for expectations in
the immediate future.
First, experimental milestones of the past five years:
\begin{itemize}
	\item  First estimates of $\Delta g(x,Q^{2})$ from evolution.
	\item  First good look at $g_{2}(x,Q^{2})$ from SLAC\null. 
	\item  First measurement of $\mu_{s}\equiv\langle\half
	\vec r\times s^{\dagger}\vec \alpha s\rangle$ from SAMPLE.
	\item  First fragmentation asymmetry measurements from Hermes.
	\item  Commissioning of the polarized $pp$ component of 
	RHIC\null.
\end{itemize}
Next, theory milestones of the past five years:
\begin{itemize}
	\item  Theory of $\Delta g$ measurements:  Via evolution, via $\bar 
	c c$ production, via ${\vec p}_{\parallel}{\vec p}_{\parallel}\to \gamma 
	\hbox { jet } X$.
	\item Development of the physics program for 
    	$\overrightarrow{\rm RHIC}$.
	\item Off-forward parton distributions and their possible 
	measurement in deeply virtual Compton scattering (DVCS).
	\item The theory of the nucleon's angular momentum.
	\item The theory of transversity and proposals to measure it.
	\item The classification of spin and transverse momentum effects in 
	distribution and fragmentation processes.
\end{itemize}
And finally, prospects for the near  future:
\begin{itemize}
	\item Direct measurements of $\Delta g$.
	\item First measurements of transversity.
	\item First measurements of polarization-dependent fragmentation 
	functions.
	\item Study of the inclusive/exclusive connection (i.e., higher twist),
	photoproduction, and DVCS at JLab.
	\item Flavor separation of the quark spin distributions.
	\item High quality measurements of $\Delta q(x,Q^{2})$  at
	very low and very high $x$.
\end{itemize}

Clearly this field -- the study 
of QCD confinement dynamics using polarized probes -- requires more than a
single facility.  Low $Q^{2}$ and low energy are  needed for DVCS and for
studies of higher twist.  High $Q^{2}$ is  needed to study $\Delta g$ via
evolution.  High energy is necessary to  create the phase space for complex
final states such as
$\bar c c$  studies of $\Delta g$ and multijet final states.  Both polarized 
lepton beams and polarized proton beams are required.  High-density 
polarized targets are required for high-luminosity studies of $g_{2}$ at 
SLAC and extraction of neutron 
distributions from polarized deuterium and $^{3}$He scattering data.

While we enthuse about the particular subject of this 
workshop, we must remember that the field requires an opportunistic, 
even predatory mentality, ready to make use of many facilities in 
imaginative ways.

\subsection{Bjorken's Sum Rule}

Occasionally it is worth reminding ourselves what it means to
``understand'' something in QCD. In the absence of fundamental
understanding we often invoke ``effective descriptions'' based on
symmetries and low-energy expansions.  While they can be extremely
useful, we should not forget that a thorough understanding allows us
to relate phenomena at very different distance scales to one another. 
In the case case of Bjorken's sum rule, the operator product
expansion, renormalization group invariance and isospin conservation
combine to relate deep inelastic scattering at high $Q^{2}$ to the
neutron's $\beta$-decay axial charge measured at very low energy. 
Even target mass and higher twist corrections are relatively well
understood.  The present state of the sum rule is
\bea
	\int_{0}^{1}\rd x g_{1}^{ep-en}(x,Q^{2})&=&
	\frac{1}{6}\frac{g_{A}}{g_{V}}\left\{1-\frac{\alpha_{s}(Q^{2})}{\pi}
	-\frac{43}{12}\frac{\alpha_{s}^{2}(Q^{2})}{\pi^{2}}
	-20.215\frac{\alpha_{s}^{3}(Q^{2})}{\pi^{3}}\right\}\nonumber\\
	&&\quad{}+\frac{M^{2}}{Q^{2}}\int_{0}^{1}x^{2}\rd x \left\{
	\frac{2}{9}g_{1}^{ep-en}(x,Q^{2})+\frac{1}{6}
	g_{2}^{ep-en}(x,Q^{2})\right\}\nonumber\\
	&&\qquad{}- \frac{1}{Q^{2}}\frac{4}{27}{\cal F}^{u-d}(Q^{2})
	\label{eq2}
\eea
where the three lines correspond to QCD~\cite{Larin:1991tj}, target
mass, and higher twist\cite{Shuryak:1982pi} corrections respectively; 
$g_{1}$ and $g_{2}$ are the nucleon's longitudinal and transverse
spin-dependent structure functions;  $g_{A}$ and $g_{V}$ are the neutron's
$\beta$-decay axial and vector charges.  ${\cal F}$ is a twist-4
operator matrix element with dimensions of $[\hbox{mass}]^{2}$, which
measures a quark-gluon correlation within the nucleon,
\be
	{\cal F}^{u}(Q^{2})s^{\alpha}=\half\langle PS|\left.g\bar
u\widetilde F^{\alpha\lambda}\gamma_{\lambda}u\right|_{Q^{2}}|PS\rangle
	\la{eq3}
\ee
where $g$ is the QCD coupling, $\widetilde F$ is the dual 
gluon field strength, and $|_{Q^{2}}$ denotes the operator 
renormalization point.

\begin{figure}[ht]
$$
\BoxedEPSF{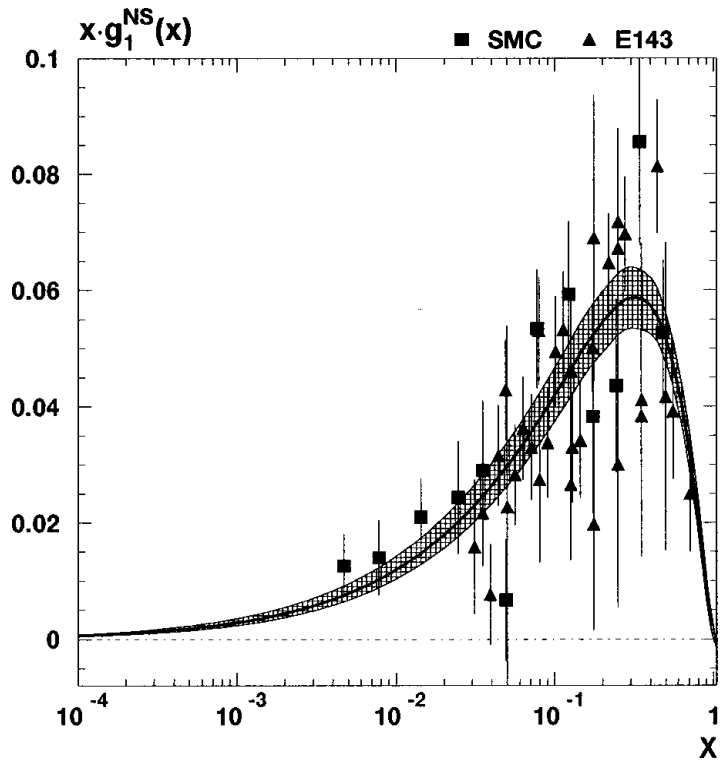 scaled 1200}   
$$
\caption{SMC analysis of data relevant to the Bjorken sum rule.}
\label{BjSR2}
\end{figure}

The most thorough analysis of the Bj sum rule I know of is one
presented by SMC in 1998~\cite{Adeva:1998vw}. Their theoretical
evaluation gives
\be
	\int_{0}^{1}\rd x g_{1}^{ep-en}(x,Q^{2})|_{\mathrm{theory}}=0.181\pm
0.003
	\la{Bjtheory}
\ee
at $Q^{2}=5$ GeV$^{2}$. Experiment is not yet able
to reach this level of accuracy.  The latest data relevant to the Bj
sum rule is shown in Fig.~\ref{BjSR2}.  The value extracted by the SMC
is
\be
	\int_{0}^{1}\rd x g_{1}^{ep-en}(x,Q^{2})|_{\mathrm expt}=0.174 
	\pm 0.05
	\begin{array}{ll}+0.011\\ -0.009\end{array}
	\begin{array}{ll}+0.021\\ -0.006\end{array}
	\la{Bjexpt}
\ee
at $Q^{2}=5$ GeV$^{2}$, and the errors are statistical, systematic,
and ``theoretical'' (e.g.,  generated by running the data to a common
$Q^{2}$), respectively~\cite{Adeva:1998vw}. Further accuracy is
necessary to confirm the target mass corrections and extract the twist-four
contribution.

\subsection{Quark and gluon distributions in the nucleon}

No overview of the nucleon's spin structure is complete without a 
survey of the polarized quark and gluon distributions in the nucleon.  
These helicity-weighted momentum distributions are the most precise 
and interpretable information we have about the spin substructure of 
a hadron.  The distributions are usually defined in terms of flavor-SU(3)
structure,
\bea
	\hbox{Singlet:}\quad\Delta\Sigma &=& \Delta U +\Delta D +\Delta 
	S\nonumber\\
	\hbox{Nonsinglet, isovector:}\quad\Delta q_{3}&=& \Delta U -\Delta D \nonumber\\
	\hbox{Nonsinglet, hypercharge:}\quad\Delta q_{8} &=& \Delta U +\Delta D -2\Delta 
	S 
	\la{qg-dist}
\eea
where $\Delta Q\equiv q^{\uparrow}(x,Q^{2})+\bar q^{\uparrow}(x,Q^{2})
-q^{\downarrow}(x,Q^{2})-\bar q^{\downarrow}(x,Q^{2})$.  Experimenters 
seem to prefer nonsinglet distributions specialized to the proton and 
neutron individually,
\bea
	\hbox{Proton nonsinglet:}\quad\Delta q_{NS}(p)&=& \Delta U -\half\Delta D
	-\half\Delta 
	S\nonumber\\
	\hbox{Neutron nonsinglet:}\quad\Delta q_{NS}(n)&=& \Delta D -\half\Delta 
	U-\half\Delta S 
	\la{qg-nonsinglet}
\eea
so that
\bea
	g_{1}^{p} &=& \fract{2}{9}\Delta\Sigma +\fract{2}{9}\Delta q_{NS}(p)
	\nonumber\\
	g_{1}^{n} &=& \fract{2}{9}\Delta\Sigma +\fract{2}{9}\Delta q_{NS}(n)\ .
	\la{q-pn}
\eea

Since the integrated quark spin accounts for only about 30\% of the
nucleon's spin, it is extremely interesting to know whether the
integrated gluon spin in the nucleon is large.  Of course the
polarized gluon distribution, $\Delta g(x,Q^{2})$, cannot be measured
directly in deep inelastic scattering because gluons do not couple to
the electromagnetic current.  Instead, $\Delta g$ is inferred from the
QCD evolution of the quark distributions.  (See
Ref.~\cite{Adeva:1998vw} for details of the process and references to
the original literature.)  However, evolution of imprecise data only
constrains a few low moments of $\Delta g$ and gives only crude
information on global characteristics such as the existence and number
of nodes.  It is clear that $\Delta g$ must be measured directly
elsewhere.

\begin{figure}[ht]
$$
\BoxedEPSF{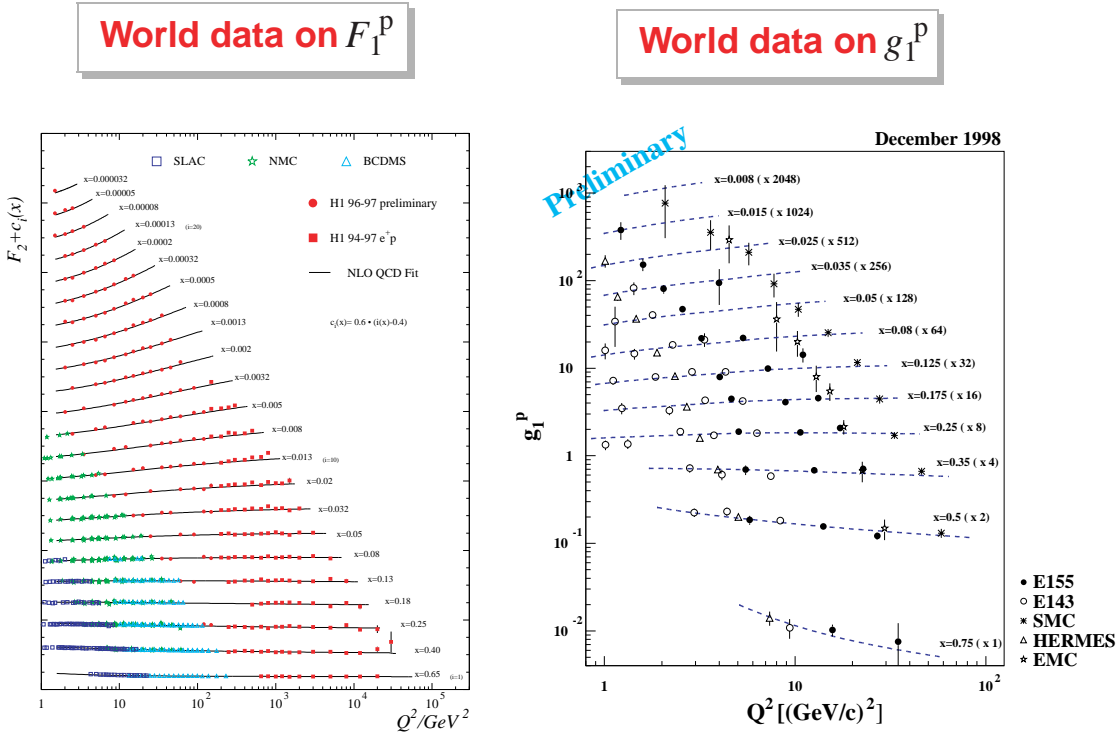 scaled 550}   
$$\smallskip
\caption{World data on spin-average and spin-dependent structure
functions~\protect\cite{Makins}.}
\label{Makins3}
\end{figure}

\begin{figure}[ht]
$$
\BoxedEPSF{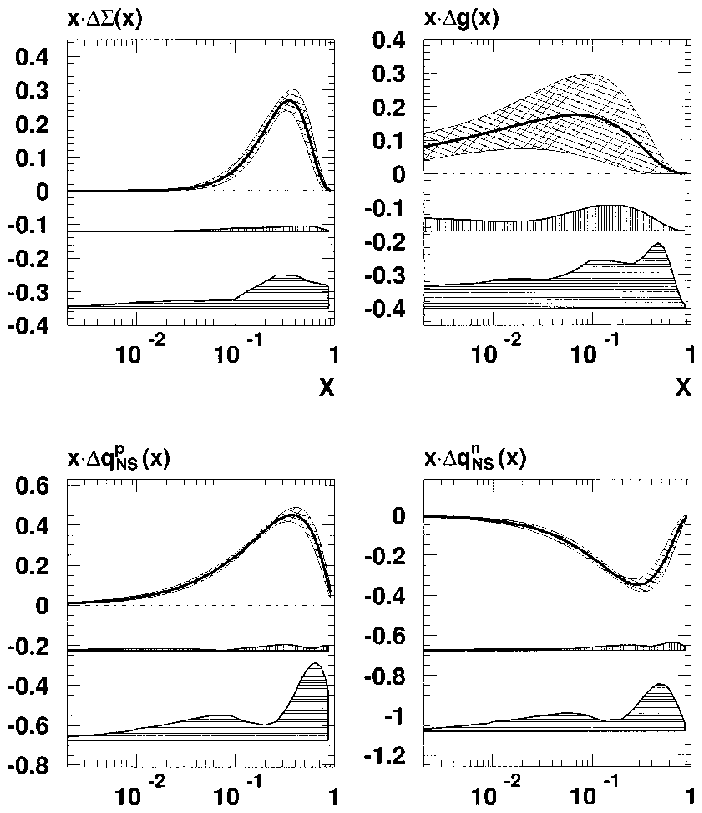 scaled 1600}  
$$\medskip
\caption{Polarized quark and gluon distribution functions. The upper figures
show the distribution with a statistical error bound. The lower figures show
estimates of systematic and theoretical uncertainties, respectively.}
\label{SMC4}
\end{figure}

That said, the world's data on polarized structure functions is 
summarized in Figs.~\ref{Makins3} and \ref{SMC4}.  Fig.~\ref{Makins3} is 
taken from Naomi Makins's talk at DIS2000 and presents the world's 
data on $g_{1}^{p}$ in the same format traditionally used for
unpolarized structure function data~\cite{Makins}. The figure highlights
the tremendous progress of the past decade as well as the need for
much better data if our knowledge of polarized distributions would
aspire to the same accuracy as unpolarized distributions.  Note, in
particular, that the entire kinematic domain over which $g_{1}$ has
been measured would fit into the lower left-hand corner of the $F_{2}$
figure.  Fig.~\ref{SMC4} shows the quark and gluon distributions
extracted from the world's data by
SMC, together with estimates of systematic and theoretical
uncertainties~\cite{Adeva:1998vw}. While the information on quark
distributions is fairly precise, it is clear that we know very little
about the distribution of polarized gluons in the nucleon.


\section{Is there an ``Angular Momentum Sum Rule'' and is it 
experimentally testable?}
\label{section2}

It has been clear for years that in some sense the nucleon's spin
(projected along an axis) can be written as a sum of contributions
from quark and gluon spin and orbital angular
momentum~\cite{Jaffe:1990jz},
\begin{equation}
	\half = \half\Delta\Sigma +\Delta g +L_{q}+L_{g}
	\la{angmom}
\end{equation}
but the interpretation and usefulness of such a relation has only 
recently been clarified.  The principal issues are 
\bit
	\item  Are the terms separately gauge-invariant?
	\item Are they interaction-dependent?
	\item Is each separately measurable?
   	\item Is each related to an integral  over a parton $x$-distribution?
\eit
I believe we can now answer these questions, but the answers are not 
what we would like. 

I want to distinguish between two different kinds of relations
with the form of Eq.~\ref{angmom}.  A ``classic'' sum rule expresses
the expectation value of a local operator in a state as an {\it
integral\/} (or sum) over a distribution measured in an inelastic
production process involving the same state.  This is the traditional
definition of a ``sum rule'', dating back to the Thomas-Reiche-Kuhn
sum rule of atomic spectroscopy. All the familiar sum rules
of deep inelastic scattering -- Bjorken's, Gross \& Llewellyn-Smith's,
etc.\  -- are this type of relation.  They are even more
powerful because the distribution that is integrated has a simple,
heuristic interpretation as the momentum (Bjorken-$x$) distribution of
the observable associated with the local operator.  The ``spin sum
rule'' gives a typical example:
\begin{eqnarray}
	\langle P,S|\left.\bar q_{a} \gamma^{\mu}\gamma_{5}
	q_{a}\right|_{Q^{2}}|P,S\rangle\Bigm/S^{\mu} &\equiv&
	\Delta q_{a}(Q^{2}) 	\label{eq2.0}
\\
	&=& \int_{0}^{1}\rd x \left\{q_{a\uparrow}(x,Q^{2})
	+\bar q_{a\uparrow}(x,Q^{2})-q_{a\downarrow}(x,Q^{2})
	-\bar q_{a\downarrow}(x,Q^{2})\right\} \nonumber
\end{eqnarray}
The left-hand side can be measured in $\beta$-decay or other
electroweak processes.  The right-hand side can be measured in deep
inelastic scattering of polarized leptons from polarized targets.  The
meaning of the sum rule is clear because the local operator, $\bar
q_{a} \gamma^{\mu}\gamma_{5} q_{a}$, is the generator of the  
internal rotations (the ``spin'') of the quark field in QCD.  The sum 
rule says that the quark's contribution to the nucleon's spin is the 
integral over a spin-weighted momentum distribution of the quarks.

Another, less powerful, but still interesting type of relation --
sometimes called a sum rule -- arises simply because an operator can
be written as the sum of two (or more) other operators, $\Theta =
\Theta_{1}+\Theta_{2}$.  If the expectation values of all three
operators can be measured, then this relation, and the assumptions
underlying it, can be tested.  Such a relation exists for the
contributions to the nucleon's angular
momentum~\cite{Jaffe:1990jz,Ji:1997ek},
\begin{equation}
	\half = \hat L_{q} + \half\Sigma + \hat J_{g}
	\label{eq2.01}
\end{equation}
where the three terms are {\it roughly\/} the quark orbital angular
momentum, the quark spin, and the total angular momentum on the
gluons.  Ji has shown how, in principle, to measure the various terms in
this relation~\cite{Ji:1997ek}.  

A sum rule of the classic type also exists for the contributions to
the nucleon's angular momentum,
\cite{Harindranath:1999ve,Hagler:1998kg,Bashinsky:1998if}
\begin{equation}
	\half = \int_{0}^{1}\rd x \left\{L_{q}(x,Q^{2}) + \half \Delta q(x,Q^{2})
	+ L_{g}(x,Q^{2}) + \Delta g(x, Q^{2})\right\}
	\label{eq2.02}
\end{equation}
where the four terms are {\it precisely\/} the $x$-distributions of
the quark orbital angular momentum, quark spin, gluon orbital angular
momentum, and gluon spin.  However, it appears that the
distributions $L_{q}(x,Q^{2})$ and $L_{g}(x,Q^{2})$ are not
experimentally accessible.  So the value of the sum rule is obscure.

Before exploring these relations for the angular momentum in more
depth, let's examine the simpler and well-understood case of energy and 
momentum.

\subsection{Sum rules for energy and momentum}

One hears a lot about the ``momentum sum rule'' in QCD, but nothing
about an ``energy sum rule''.  The reasons are quite instructive. 
Energy and momentum are described by the rank-two, symmetric
energy-momentum tensor, $T^{\mu\nu}$,
\begin{equation}
	T^{\mu\nu} = \frac{i}{4}\bar q(\gamma^{\mu}D^{\nu}+\gamma^{\nu}
	D^{\mu})q +\hbox{h.c.}+ 
	\Tr(F^{\mu\alpha}F_{\alpha}^{\nu} - 
	\fract{1}{4}g^{\mu\nu}F^{2}) 
	\label{eq2.1}
\end{equation}
where $D^{\mu}$ and $F^{\mu\nu}$ are the gauge covariant derivative 
and gluon field strength, both matrices in the fundamental representation
of SU(3). [$T^{\mu\nu}$ is ambiguous up to certain total 
derivatives, but these do not change the arguments presented here.]  

The energy density is given by $T^{00}$,
\begin{equation}
	{\cal E} \equiv T^{00} = \half q^{\dagger}(-i\vec\alpha\cdot\vec D
	+\beta m)q + \hbox{h.c.} + \Tr(\vec E^{2}+ \vec B^{2})\, .
	\label{eq2.2}
\end{equation}
The expectation value of $T^{00}$ is normalized,
\begin{equation}
	\langle P | T^{00} | P \rangle = 2E^{2}\, ,
	\label{eq2.3}
\end{equation}
because $|P\rangle$ is an eigenstate of the Hamiltonian, $\int \rd{^3 x}
T^{00}(x)|P\rangle = E|P\rangle$.  This is a good start towards a sum
rule.  However there is no useful sum rule because there is no way to
write any of the terms in \eq{eq2.2} as an integral over inelastic
production data.  This is not obvious, but the appearance of terms in
${\cal E}$ that are order cubic and higher in the canonical fields is
a bad sign. The first term in ${\cal E}$ includes $\bar
q q g$ coupling, and $\vec E^{2}+\vec B^{2}$ involves terms cubic and
quartic in the gluon vector potentials $\vec A$. The parton distributions of deep inelastic scattering
(DIS) come from operators quadratic in the ``good'' light-cone
components of the quark and gluon fields, $q_{+}$ and $\vec
A_{\perp}$~\cite{Jaffe:zw}. 

In contrast there is a classic, deep-inelastic sum rule for $P^{+}$, 
where $P^{+}=\frac{1}{\sqrt{2}}(P^{0}+P^{3})$, and the 3-direction is 
singled out by the gauge choice $A^{+}=0$.  $T^{++}$ is normalized 
much like $T^{00}$,
\begin{equation}
	\langle P |T^{++}|P\rangle = 2{P^{+}}^{2}.
	\label{eq2.4}
\end{equation}
Unlike $T^{00}$, $T^{++}$ simplifies dramatically in $A^{+}=0$ gauge 
because of the simplification of $D^{+}$ and $F^{+\alpha}$,
\begin{eqnarray}
	D^{+}&=&\partial^{+}-igA^{+} \to \partial^{+}\nonumber\\
	F^{+\alpha}&=&\partial^{+}A^{\alpha}-\partial^{\alpha}A^{+}
	+g[A^{+},A^{\alpha}] \to \partial^{+}A^{\alpha}\, .\label{eq2.5}
\end{eqnarray}
As a result $T^{++}$ is quadratic in the fundamental dynamical
variables, $q_{+}$ and $\vec A_{\perp}$ and all interactions disappear,
\begin{equation}
	T^{++} = iq_{+}^{\dagger}\partial^{+} q_{+} + 
	\Tr(\partial^{+}\vec A_{\perp})^{2}\  .
	\label{eq2.6}
\end{equation}
The two terms give the contributions of quarks and gluons respectively 
to the total $P^{+}$.  It is straightforward to relate each to 
an integral over a positive definite parton ``momentum'' distribution,
\begin{eqnarray}
	iq_{+}^{\dagger}\partial^{+} q &\to &\int \rd x x q(x)\nonumber\\
	(\partial^{+}\vec A_{\perp})^{2} &\to &\int \rd x x g(x)\, 
	\label{eq2.7}
\end{eqnarray}
in which the parton probability density is weighted by the 
observable (in this case $x$) appropriate to the sum rule.
Keeping track of renormalization scale dependence and kinematic
factors of $P^{+}$, one obtains the standard ``Momentum'' sum rule,
\begin{equation}
	1 = \int_{0}^{1} \rd x x \left\{q(x,Q^{2}) + g(x,Q^{2})\right\}
	\label{eq2.8}
\end{equation}

The lessons learned from this exercise generalize to the more 
difficult case of angular momentum:
\begin{itemize}
	\item The time components of the tensor densities associated with
	space-time symmetries do not yield classic sum rules. 
	Interactions do not drop out.  They yield relations that are
	difficult to interpret because quark and gluon contributions do
	not separate.  Individual terms are not related to integrals over
	parton distributions.

	\item  The $+$-components of the same tensor densities do yield 
	useful sum rules, which have a parton interpretation in $A^{+}=0$ 
	gauge.  Interactions drop out.  Each term can be represented as an 
	integral over a parton distribution weighted by the 
	appropriate observable quantity. 
\end{itemize}

\subsection{Sum rules for angular momentum}

The situation for angular momentum is not satisfactory.  The
time-component analysis yields a relation, some of whose ingredients
can be measured (in principle) in deeply virtual Compton scattering. 
But it has no place for a separately gauge-invariant gluon spin and
orbital angular momentum, no clean separation between quark and gluon
contributions, and no relation to quark or gluon $x$ distributions. 
The $+$-component analysis yields a classic sum rule with separate
quark and gluon spin and orbital angular momentum contributions, each
gauge invariant, each related to a parton distribution, and each free
from interaction terms.  Unfortunately, there does not seem to be a
way to measure the terms in this otherwise perfectly satisfactory 
sum rule.

The tensor density associated with rotations and boosts is a three
component tensor antisymmetric in the last two indices, $M^{\mnl}$. 
To extract a sum rule, we polarize the nucleon along the 3-direction
in its rest frame and set $\nu=1,\lambda=2$ in order to select
rotations about this direction.  The matrix elements of $M^{012}$ and
$M^{+12}$ are both normalized in terms of the nucleon's momentum 
($P^{\mu}=(M,0,0,0)$) and spin ($S^{\mu}=(0,0,0,M)$)~\cite{Jaffe:1990jz}.

First consider the time component, $M^{012}$,
\begin{equation}
	M^{012} = \frac{i}{2}q^{\dagger}(\vec x \times \vec D)^{3}q
	+\half q^{\dagger}\sigma^{3} q +2 \Tr E^{j}
	(\vec x \times \vec D)^{3} A^{j} + \Tr  (\vec E\times
	\vec A)^{3}\, .
	\label{eq2.9}
\end{equation}
The four terms look like the generators of rotations (about the
3-axis) for quark orbital, quark spin, gluon orbital, and gluon spin
angular momentum respectively.  Taking the matrix element in a nucleon
state at rest one obtains,
\begin{equation}
	\half = \hat L_{q} + \half\Sigma + \hat L_{g} + \Delta \hat g
	\label{eq2.10}
\end{equation}
There are problems, however.  There are no parton representations for
$\hat L_{g}$, $\hat L_{q}$, or $\Delta\hat g$, so it is not a sum rule
in the classic sense.  $\Sigma$ is the integral of the helicity
weighted quark distribution, but $\Delta\hat g$ is not the integral of
the helicity weighted gluon distribution.  Interactions prevent a
clean separation into quark and gluon contributions as they did for
$T^{00}$.  And worse still, $\hat L_{g}$ and $\Delta\hat g$ are not
separately gauge invariant, so only the sum $\hat J_{g}=\hat
L_{g}+\Delta\hat g$ is physically meaningful.

The most important feature of the relation, \eq{eq2.10}, is the result
derived by Ji, that 
\begin{equation}
	\hat J_{q}=\hat L_{q} +\half\Sigma
	\label{Ji}
\end{equation}
and $\hat J_{g}$ can, in principle, be measured in deeply virtual
Compton scattering~\cite{Ji:1997ek}. In practice, $\hat L_{q}$ may be
measurable, but $\hat J_{g}$ can only be obtained by $Q^{2}$ evolution
of $\hat J_{q}$, which seems beyond experimental attack for the
foreseeable future.  Without a handle on $\hat J_{g}$ and given the
ambiguity in the definition of $\hat L_{q}$ (see below), the
usefulness of Eq.~\ref{Ji} is unclear.

Turning to the $+$-component sum rule, we find a much simpler form,
\begin{equation}
	M^{+12} = \fract{1}{2} 
	q^{\dagger}_{+}(\vec x\times\vec i\partial)^{3}q_{+}
	+\half q_{+}^{\dagger}\gamma_{5} q_{+}
	+ 2\Tr F^{+j}(\vec x\times i \vec\partial)A^{j}
	+\Tr  \epsilon^{+-ij}F^{+i}A^{j}
	\label{eq2.11}
\end{equation}
in $A^{+}=0$ gauge. [This gauge condition must be supplemented
by the additional condition that the gauge fields vanish fast enough
at infinity.] The four terms in $M^{+12}$ correspond respectively to
quark orbital angular momentum, quark spin, gluon orbital angular
momentum, and gluon spin, all about the 3-axis.  Each is separately
gauge invariant\footnote{Note, however, that in any gauge other than
$A^{+}=0$, the operators are nonlocal and appear to be interaction
dependent.  The same happens to the simple operators involved in the
momentum sum rule, \eq{eq2.6}.} and involves only the ``good'', i.e.,
dynamically independent, degrees of freedom, $q_{+}$ and $\vec A_{\perp}$. 
Each is a generator of the appropriate symmetry transformation in light-front
field theory.  The resulting sum rule,
\begin{equation}
	\half = L_{q} + \half \Sigma + L_{g} + \Delta g
	\label{eq2.12}
\end{equation}
is a classic deep inelastic sum rule.  It can be written as an 
integral over $x$-distributions
\begin{equation}
	\half = \int_{0}^{1}\rd x\left\{L_{q}(x,Q^{2}) + \half \delta q(x,Q^{2})
	+ L_{g}(x,Q^{2}) + \Delta g(x, Q^{2})\right\}
	\label{eq2.13}
\end{equation}
where each term is an interaction independent, gauge invariant, 
integral over a partonic density associated with the appropriate 
symmetry 
generator~\cite{Hagler:1998kg,Harindranath:1999ve,Bashinsky:1998if}.

Satisfying though \eqs{eq2.12}{eq2.13} may be from a theoretical point 
of view.  They are quite useless unless someone finds a way to measure 
the two new terms $L_{q}$ and $L_{g}$.

\section{Transversity}

One of the major accomplishments of the recent renaissance in QCD spin
physics has been the rediscovery and exploration of the quark {\it
transversity distribution}.  First mentioned by Ralston and Soper in
1979 in their treatment of Drell-Yan $\mu$-pair production by
transversely polarized protons~\cite{Ralston:1979ys}, the transversity
was not recognized as a major component in the description of the
nucleon's spin until the early
1990s~\cite{Artru:1990zv,Jaffe:1991kp,Cortes:1992ja,Jaffe:zw}.

The transversity can be interpreted in parton language as follows:
consider a nucleon moving with (infinite) momentum in the $\hat
e_{3}$-direction, but polarized along one of the directions transverse
to $\hat e_{3}$.  $\delta q_{a}(x,Q^{2})$ counts the quarks of flavor
$a$ and  momentum fraction $x$ with their spin parallel the spin of a
nucleon minus the number antiparallel.  If quarks moved
nonrelativistically in the nucleon, $\delta q$ and $\Delta q$ would
be identical, since rotations and Euclidean boosts commute and a
series of boosts and rotations can convert a longitudinally polarized
nucleon into a transversely polarized nucleon at infinite momentum. 
So the difference between the transversity and helicity distributions
reflects the relativistic character of quark motion in the nucleon. 
There are other important differences between transversity and
helicity.  For example, quark and gluon helicity distributions
($\Delta q$ and $\Delta g$) mix under $Q^{2}$-evolution.  There is no
analog of gluon transversity in the nucleon, so $\delta q$ evolves
without mixing, like a nonsinglet distribution function.  The lowest
moment of the transversity is proportional to the nucleon matrix
element of the tensor charge, $\bar q i\sigma^{0i}\gamma_{5}q$, which
couples only to valence quarks (it is $C$-odd).  Not coupling to glue or
$\bar q q$ pairs, the tensor charge promises to be more
 quark-model--like than the axial charge and should be an 
interesting contrast.

\begin{figure}[ht]
 \begin{center}
 \BoxedEPSF{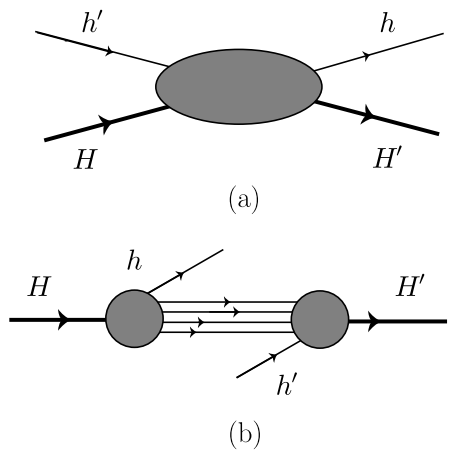 scaled 600} 
 \end{center}
 \caption{Quark hadron forward scattering.  Quark helicities are 
  labeled $h$ and $h'$; hadron helicities are $H$ and $H'$. (a)~Full 
  scattering amplitude; (b)~u-channel discontinuity, which gives the 
  quark distribution function in DIS\null.}
 \label{fig1}
 \end{figure}

We now know that the transversity, $\delta q(x,Q^{2})$, together with
the unpolarized distribution, $q(x,Q^{2})$, and the helicity
distribution, $\Delta q(x,Q^{2})$, are required to give a complete
description of the quark spin in the nucleon at leading twist.  An
equation tells this story clearly:
\begin{equation}
{\cal A}(x,Q^2) = \half q(x,Q^2)~I\otimes I + \half \Delta
q(x,Q^2)~\sigma_3 \otimes 
\sigma_3+\half  \delta q(x,Q^2)~
\left(\sigma_+\otimes\sigma_-+\sigma_-\otimes\sigma_+\right)\,. 
\label{symmetry}
\end{equation}
Here, ${\cal A}$ is the quark distribution in a nucleon as a density
matrix in both the quark and nucleon helicities (hence the external
product of two Pauli matrices in each term), diagrammatically 
equivalent to the lower part of the handbag diagram shown in 
Fig.~\ref{fig1}a;  
$q$~governs spin average physics, $\Delta q$ governs helicity
dependence, and $\delta q$ governs helicity flip -- or transverse
polarization -- physics.

In terms of the helicity amplitude ${\cal A}_{Hh,H'h'}$ in
Fig.~\ref{fig1}b, the transversity is given by ${\cal A}_{++,--}$,
corresponding to quark and nucleon helicity flip.  The spin average ($q$)
and helicity ($\Delta q$) distributions involve ${\cal A}_{++,++}$,
${\cal A}_{+-,+-}$, which preserve quark helicity.  The connection
between transverse spin and helicity flip is a consequence of simple
quantum mechanics.  The two states of transverse polarization can be
written as superpositions of helicity eigenstates: $|{\perp \atop
\top}\rangle= \frac{1}{\sqrt{2}}(|+\rangle\pm|-\rangle)$; the cross
section with transverse polarization has the form
$d\sigma_{\perp\atop\top}\propto \langle {\perp\atop\top}|\ldots|
{\perp\atop\top}\rangle$; so the difference of cross sections is
proportional to helicity flip, $d\sigma_{\perp}-d\sigma_{\top} \propto
\langle +|\ldots|-\rangle + \langle -|\ldots|+\rangle$.  At leading twist,
quark helicity and chirality are identical.  For this
reason, the transversity distribution are called ``chiral-odd'', in
contrast to the ``chiral-even'' distributions, $q$ and $\Delta q$.

Quark chirality is conserved at all QCD and electroweak vertices;
however, quark chirality can flip in distribution and fragmentation
functions because they probe the soft regime where chiral symmetry is
dynamically broken in QCD. This is another reason to be interested in
transversity -- it probes dynamical chiral symmetry breaking, an
incompletely understood aspect of QCD.

\begin{figure}[ht]
$$
\BoxedEPSF{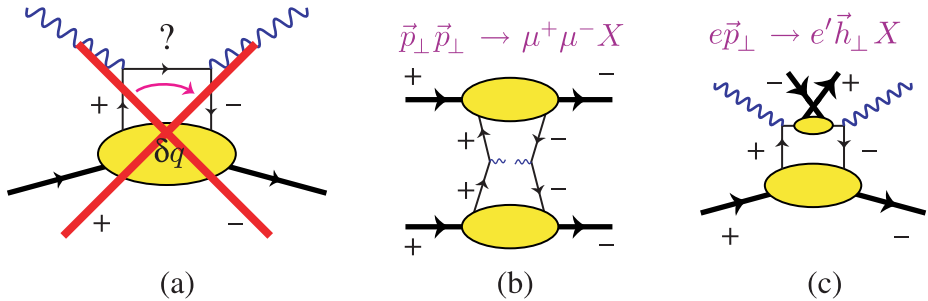 scaled 1000}  
$$
\caption{Deep inelastic processes relevant to transversity.}
\label{decouple10}
\end{figure}

Because all hard QCD and electroweak processes preserve chirality,
transversity is difficult to measure.  It decouples from inclusive DIS
and most other familiar deep inelastic processes.  The argument is
made graphically in Fig.~\ref{decouple10}.
In order to access transversity some second soft process must flip the
quark chirality a second time.  The classic example, where
transversity was discovered by Ralston and Soper, is transversely
polarized Drell-Yan production of muon pairs: $\vec p_{\perp}\vec
p_{\perp}\to \mu^{+}\mu^{-}X$, which is shown diagrammatically in
Fig.~\ref{decouple10}(b).  Chirality is flipped in both soft distribution functions and the cross
section is proportional to $\delta q(x_{1},Q^{2})\times\delta\bar
q(x_{2},Q^{2})$.

Transversity would not decouple from deep inelastic scattering if some
electroweak vertex would flip chirality.  Unfortunately (and
accidentally from the point of view of QCD) all photon, $W^{\pm}$ and
$Z^{0}$ couplings all preserve chirality.  Quark-Higgs couplings
violate chirality but are too weak to be of interest.  Quark mass
insertions flip chirality, and indeed a
careful analysis reveals effects proportional to $m\delta
q(x,Q^{2})/\sqrt{Q^{2}}$ in inclusive DIS with a transversely
polarized target.  However the $u$, $d$, and $s$ quarks, which are
common in the nucleon, are too light to give significant sensitivity
to $\delta q$.

What is needed is an insertion that flips chirality without
introducing a $1/\sqrt{Q^{2}}$ suppression.  A generic example is
shown in Fig.~\ref{decouple10}(c).  Much interest has been generated recently
by the observation of an asymmetry at Hermes that can be interpreted
as evidence for a chirality-flipping fragmentation function that
couples to the nucleon's transversity.  It corresponds to a particular
instance of Fig.~\ref{decouple10}(c).  If this effect is confirmed it suggests
a bright future for transversity measurements at the next generation
of polarized lepton-hadron facilities.

\section{Fragmentation and spin: The Hermes asymmetry\protect\newline
and beyond}

To my mind, the single most interesting development in QCD spin physics
over the past two years is the azimuthal asymmetry in pion
electroproduction reported by Hermes~\cite{Airapetian:2000tv}. It is
interesting both in itself and as an emblem of a new class of spin
measurements involving spin-dependent fragmentation processes, which
act as filters for exotic parton distribution functions like
transversity.

\subsection{Fragmentation functions as probes of unstable hadrons}

Fragmentation functions allow us to access and explore the spin
structure of unstable hadrons, which cannot be used as targets for
deep inelastic scattering.  These include the $\rho$, $\omega$, 
and $\phi$ mesons, and hyperons like the $\Lambda$ and $\Sigma$.
Let me give three examples:

\subsubsection{The tensor fragmentation function of the $\rho$}

When a quark of helicity $h=\pm\half$ fragments collinearity into a
$\rho$ of helicity $H=1$, $0$, or  $-1$, there are many
fragmentation functions, ${\cal F}_{Hh,H'h'}$, in analogy to ${\cal
A}_{Hh,H'h'}$ discussed in the previous section.  If we consider
fragmentation of helicity eigenstates, then the fragmentation
functions can be labelled by the quark helicity $h$ and the $\rho$
helicity $H$ corresponding to $q_{h}\to\rho_{H}$.  Parity relates
three pairs, e.g.,  $q_{\frac12}\to\rho_{1}=q_{-\frac12}\to\rho_{-1}$,
leaving three independent combinations.  These can be classified as
the spin average: $q\to\rho$; the helicity difference:
$(q_{\frac12}\to\rho_{1})- (q_{\frac12}\to\rho_{-1})$; and the tensor
fragmentation function, known as $\hat b_{\rho}$ in analogy to the
tensor {\it distribution\/} function first analyzed in connections
with the deuteron~\cite{Hoodbhoy:1989am}: $\hat
b_{\rho}=(q\to\rho_{1})+(q\to\rho_{-1})-2(q\to\rho_{0})$.  The function $\hat
b_{\rho}$ is independent of quark spin and has the simple physical
interpretation of measuring the difference between quark fragmentation
into a transverse $\rho$ compared to a longitudinal $\rho$.  The
$\pi\pi$ angular distribution in $\rho$ decay is sensitive to $\hat
b_{\rho}$, so it can be measured~\cite{Schafer:1999am}. The data are
already available.  The challenge to theorists is to make use of it.

\subsubsection{The $\rho$ double-helicity flip-fragmentation function}

\begin{figure}[ht]
$$
\BoxedEPSF{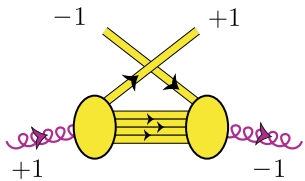 scaled 1000}  
$$
\caption{ Figure showing gluon and rho helicity labels in the double helicity 
flip case.}
\label{doubleglue}
\end{figure}

Consider the fragmentation of a gluon into a $\rho$.  In addition to 
the fragmentation functions already discussed for quarks, a 
double helicity flip fragmentation function can occur.  The process 
and the helicity labels are shown in Fig.~\ref{doubleglue}.  This 
process has a unique signature in the $\pi\pi$ angular distribution 
and no equivalent in $q\to\rho$.  So it is a special probe of gluon 
fragmentation into the $\rho$.

\subsubsection{Polarized quark $\to$ polarized $\Lambda$ fragmentation}

It should be clear that one can define longitudinal and transverse
spin dependent fragmentation functions of the $\Lambda$, schematically
$\vec q_{\parallel}\to \vec \Lambda_{\parallel}$ and $\vec
q_{\perp}\to\vec \Lambda_{\perp}$, in direct analogy to the quark
helicity and transversity distributions in a target $\Lambda$.  Since
the $\Lambda\to p \pi$ decay is self-analyzing, it is relatively easy
to measure the spin of the $\Lambda$.  By selecting $\Lambda$'s
produced in the current fragmentation region one can hope to isolate
the fragmentation process $q\to\Lambda$.  Having measured the quark
spin structure of the nucleon, we can use flavor-SU(3) to estimate the
way quark spins are distributed in the
$\Lambda$~\cite{Burkardt:1993zh}. However we do not know if this
information is reflected in the fragmentation process $q\to\Lambda$. 
Once again the challenge is to theorists to learn how to interpret
fragmentation functions in a heuristic way analogous to the quark
parton model of distribution functions.

\subsection{Fragmentation as a filter for novel distribution functions}

Even if we do not know how to interpret fragmentation functions, we
can use them as filters, to select parton distribution functions that either
decouple from or are hard to extract from completely inclusive DIS\null. The
simplest and best known example is the use of meson flavor to tag strange
versus nonstrange quark distribution functions.  This analysis has been
developed to a high level of sophistication by the Hermes collaboration who
use the felicitous term ``purity'' to denote the propensity for strange quarks
to fragment to strange mesons and so forth~\cite{Niczyporuk:1998uz}.
They identify ``favored'' fragmentation processes like $u\to\pi^{+}$
and ``disfavored'' processes like $u\to\pi^{-}$ and set up a transfer
matrix formalism to give a complete characterization of $eN\to e'
(\pi,K,\eta) X$.  The interest is not principally in the various
fragmentation functions, but instead to use them as filters for
specific quark (and antiquark) distribution functions.  Hermes and
Compass hope to use these methods to extract the polarized antiquark
distributions in the nucleon, $\Delta\bar u(x,Q^{2})$, $\Delta\bar
d(x,Q^{2})$, and $\Delta\bar s(x,Q^{2})$.  Their competition in this
pursuit comes from $\overrightarrow{\rm RHIC}$, where $W^{\pm}$
production asymmetries can be used to trigger on specific quark
flavors and extract $\Delta\bar u$ and $\Delta\bar d$.

A more complex, and potentially much more interesting example is the
use of a helicity flip fragmentation function to select the quark
transversity distribution.  As shown in Fig.~\ref{decouple10}(c), by
interposing a helicity flip fragmentation function on the struck quark
line in DIS, it is possible to access the transversity.  What is
needed is a {\it twist-two}, chiral-odd fragmentation function.  There
are several candidates:
\begin{itemize}
	\item $\delta \hat q_{a}(z,Q^{2})$, the transverse, spin-dependent
	fragmentation function.  This is the analog in fragmentation of
	transversity, and describes the fragmentation of a transversely
	polarized quark into a transversely polarized hadron with momentum
     	fraction $z$~\cite{Jaffe:1993xb,Boer:2000xx}. To access $\delta \hat q$, it is
	necessary to measure the spin of a particle in the final state of
	DIS\null. In practice this limits the application to production of a
	$\Lambda$ hyperon -- the only particle whose spin is easy to
	measure through its parity violating decay.
	
	\item $\delta \hat q_{I}(z,m^{2},Q^{2})$, the two pion
	interference fragmentation function. 
	\cite{Collins:1994kq,Collins:1994ax,Jaffe:1998hf} This describes
	the fragmentation of a transversely polarized quark into a pair of
	pions whose orbital angular momentum is correlated with the quark
	spin.  This requires measurement of two pions in the final state. 
	It may be quite useful, especially in polarized collider
	experiments~\cite{Spin2000-Perdekamp}. I will not discuss it further
	here.
	
	\item $\hat c(z,Q^{2})$, the single particle azimuthal asymmetry
	fragmentation function.  This function, first discussed by
	Collins, et al.~\cite{Collins:1994kq}, describes the azimuthal
	distribution of pions about the axis defined by the struck quark's
	momentum in deep inelastic scattering.
\end{itemize}
All three of these fragmentation functions are chiral-odd and
therefore produce experimental signatures sensitive to the
transversity distribution in the target nucleon.  Each may play an
important role in future experiments aimed at probing the nucleon's
transversity.  Recently Hermes has announced observation of a spin
asymmetry that seems to be associated with the Collins function,
$c(z,Q^{2})$.  So although all three deserve discussion, I will spend
the rest of my time on the Collins function and the Hermes asymmetry.

\subsubsection{The Collins Fragmentation Function}

The standard description of fragmentation without polarization
requires a single fragmentation function usually called $D_{h}(z)$. 
It gives the probability that a quark will fragment into a hadron,
$h$, with longitudinal momentum fraction $z$. [For simplicity
I suppress the dependence of $D$ on the virtuality scale, $Q^{2}$ and
the quark flavor label $a$.] The transverse momentum of $h$ relative
to the quark is integrated out.  If the transverse momentum, $\vec
p_{\perp}$, is observed, then it is possible to construct
distributions weighted by geometric factors.  For instance,
\begin{eqnarray}
	c(z) &\propto & \int \rd{^2 p_{\perp}} D_{h}(z,\vec p_{\perp}) \cos\chi 
\nonumber \\ 
	\noalign {\hbox{where, for comparison,}}
	D(z) &\propto & \int \rd{^2 p_{\perp}} D_{h}(z,\vec p_{\perp})\ . 
	\label{eq3.1}
\end{eqnarray}
Here $D_{h}(z,\vec p_{\perp})$ is the probability for the quark to
fragment into hadron $h$ with momentum fraction $z$ and transverse
momentum $\vec p_{\perp}$;  $\chi$ is the angle between $\vec
p_{\perp}$ and some vector, $\vec w$, defined by the initial state. 
Since we don't know the direction of the quark's momentum exactly,
the transverse momentum of the hadron, $\vec p_{\perp}$, is
defined relative to some large, externally determined momentum, such 
as the momentum of the virtual photon, $\vec q$, in DIS\null.

\begin{figure}[htb]
$$
\BoxedEPSF{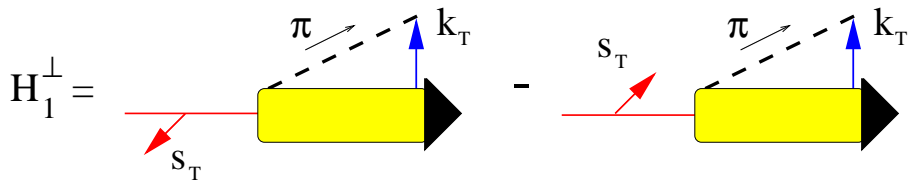 scaled 700}
$$
  \caption{\label{H1perp}The Collins effect function
  $H_1^\perp$ signals different probabilities for $q(\perp/\bot) \to
  \pi(\vec k_{\perp}) + X$.}
 \label{fig5}
 \end{figure}

How can $c(z)$ figure in deep inelastic scattering?  The trick is to
find a vector, $\vec w$, relative to which $\chi$ can be defined.  If
the target is polarized, it is possible to define $\vec w$ by taking
the cross product of the target spin, $\vec s$, with either the
initial or final electron's momentum ($\vec k$ or $\vec k'$) depending
on the circumstances.  Generically, then, the observable associated
with $c(z)$ is $\cos\chi\propto \vec k \times \vec s \cdot \vec p$,
where $\vec p$ is the momentum of the observed hadron in the final
state.  The situation is illustrated in Fig.~(\ref{fig5}) from 
Ref.~\cite{Boer:2000ya}.
This observable is even under parity (because $\vec s$ is a
pseudovector), but odd under time reversal.  This {\it does not\/}
mean that it violates time-reversal invariance.  Instead it means that
it will vanish unless there are final state interactions capable of
generating a nontrivial phase in the DIS amplitude.  This
subtlety makes it hard to find a good model to estimate $c(z)$ because
typical fragmentation models involve only tree graphs (if they involve
quantum mechanics at all!), which are real.

The Collins fragmentation function, $c(z)$, may be interesting in 
itself, but it is much more interesting because it is chiral-odd and 
combines with the transversity distribution in the initial nucleon to 
produce an experimentally observable asymmetry sensitive to 
the transversity.  Two specific cases figure in recent and 
soon-to-be-performed experiments.

\subsubsection {Single particle inclusive DIS with a transversely
polarized target: $e \vec p_{\perp}\to e' \pi X$}

If the target is transversely polarized (with respect to the initial
electron momentum, $\vec k$), then $\vec w = \vec k\times\vec s$
defines a vector normal to the plane defined by the beam and the
target spin.  The transverse momentum of the produced hadron can be
defined either with respect to the beam or the momentum transfer $\vec
q$ -- the difference in higher order in $1/Q$.  $\cos\chi$ is defined
by $\cos \chi= \vec p_{\perp}\cdot\vec w/|\vec p_{\perp}||\vec w|$. 
The kinematics are particularly simple in this case (transverse spin). 
Experimenters prefer to think of the effect in terms of the angle
($\phi$) between two planes: Plane 1 is defined by the virtual photon
and the target spin, and Plane 2 is defined by the virtual photon
and the transverse momentum of the produced hadron.  Then
$\sin\phi=\cos\chi$ and the effect is known as a ``$\sin\phi$''
asymmetry.  When the cross section is weighted by $\sin\phi$, the
result is 
\begin{equation}
	\frac{\rd{\Delta\sigma_{\perp}}}{\rd x \rd y \rd z} = 
	\frac{2\alpha^{2}}{Q^{2}} \sum_{a}e_{a}^{2}\delta 
	q_{a}(x)c_{a}(z) 
	\label{eq3.2}
\end{equation}
where $y=E-E'/E$, and $\Delta\sigma$ is the difference of cross
sections with target spin reversed.\footnote{In principle, this
reversal is superfluous because the $\sin\phi$ asymmetry must be odd
under reversal and the rest of the cross section must be even. 
However, it helps experimenters to reduce systematic errors.} This is
a leading twist effect, which scales (modulo logarithms of $Q^{2}$) in
the deep inelastic limit.  If $c(z)$ is not too small, it is will
become the ``classic'' way to measure the nucleon's transversity
distributions.

 No experimental group has yet measured hadron production in deep
 inelastic scattering from a transversely polarized target, so there
 is no data on $\Delta \sigma_{\perp}$.  Hermes at DESY intend to take
 data under these conditions in the next run.  One reason for this was
 the observation of a $\sin\phi$ asymmetry with a {\it
 longitudinally\/} polarized target that Hermes announced last
 year~\cite{Airapetian:2000tv}. It strongly suggests, but does not
 require, that $\Delta\sigma_{\perp}$ should be large.

\subsubsection {Single particle inclusive DIS with a longitudinally
polarized target: $e \vec p_{\|}\to e' \pi X$}

\begin{figure}[ht]
   \begin{center}
    \BoxedEPSF{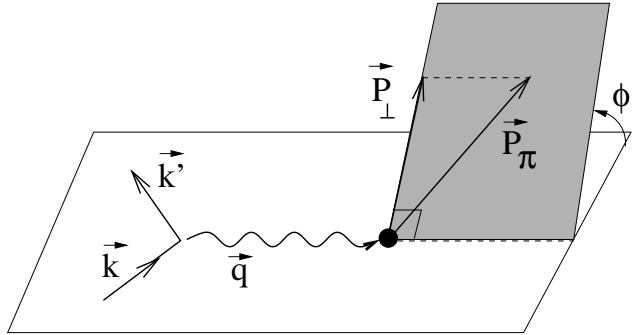 scaled 600}
    \end{center}
 \begin{minipage}[r]{\linewidth}
    \caption{Kinematic planes for pion production in semi-inclusive deep-inelastic scattering.}
  \label{fig6}
 \end{minipage}
  \end{figure}

The possibility of a $\sin\phi$ asymmetry is more subtle in this case,
and it escaped theorists' attention for a long time.  The possibility of
such an asymmetry was first pointed out in
Ref~\cite{Oganessyan:1998ma}.  As $Q^{2}$ and $\nu$ go to $\infty$,
the initial and final electrons' momenta become parallel.  If the
target spin is parallel to $\vec k$, then it is impossible to
construct a vector from $\vec k$ or $\vec k'$ and $\vec s$ in this
limit.  However, $\vec k$ and $\vec k'$ are not exactly parallel, so
$\vec s$ has a small component perpendicular to the virtual photon's
momentum, $\vec q = \vec k - \vec k'$.  The vector, $\vec w$, can be
defined as $\vec w = \vec k'\times\vec s$, and the kinematic situation
is shown in Fig.~\ref{fig6} from Ref.~\cite{Airapetian:2000tv}.
This produces an asymmetry similar to the previous case, but weighted
by $|\vec s_{\perp}| \propto 2Mx/Q$.  Because this leading (twist two)
effect is kinematically suppressed by $1/Q$, it is necessary to
consider other, twist-three, effects that might be competitive.  A
careful analysis turns up a variety of twist-three effects, leading to
a cross section of the form~\cite{Boer:2000ya,Oganessyan:1998ma},
\begin{equation}
	\frac{\rd{\Delta\sigma_{\|}}}{\rd x \rd y \rd z} = 
	\frac{2\alpha^{2}}{Q^{2}}
\frac{2Mx}{Q}\sqrt{1-y}\sum_{a}e_{a}^{2}\Bigl\{\delta 
	q_{a}(x)c_{a}(z) + \frac{2-y}{1-y} h_{La}(x)c_{a}(z)\Bigr\}
	\label{eq3.3}
\end{equation}
where $h_{L}(x)$ is a longitudinal spin-dependent, twist-three 
distribution function analogous to $g_{T}$.

\begin{figure}[ht]
 \begin{center}
 \BoxedEPSF{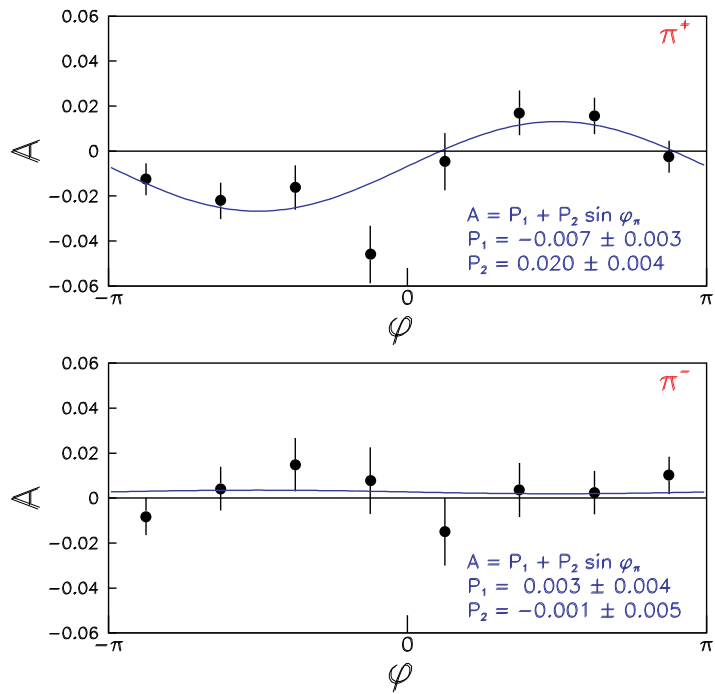 scaled 700} 
 \end{center}
  \caption{Azimuthal asymmetry ($\sin\phi$) distribution for $\pi+$ 
  and $\pi^{-}$ production with a longitudinally polarized target at 
  Hermes.}
 \label{fig7}
  \end{figure}

By far the most interesting thing about $\Delta \sigma_{\|}$ is that
Hermes has seen such an asymmetry in their $\pi^{+}$ data.  (The Hermes 
data is shown in Fig.~\ref{fig7}.)  
They see no effect in their $\pi^{-}$ data.  Because $u$ quarks
predominate in the nucleon, because $e_{u}^{2}=4e_{d}^{2}$, and
because $u\to\pi^{+}\ll u\to\pi^{-}$, they expect no signal in
$\pi^{-}$.  They have not reported on $\pi^{0}$, where an asymmetry
similar to $\pi^{+}$ would be expected.

If the Hermes result is confirmed, it demonstrates that the Collins
fragmentation function is nonzero.  Somehow the final state
interactions between the observed pion and the other fragments of the
nucleon suffice to generate a phase that survives the sum over the
other unobserved hadrons.  Whatever its origin, a nonvanishing
Collins function would be a great gift to the community interested in
the transverse spin structure of the nucleon.  It provides an
unanticipated tool for extracting the nucleon's transversity from DIS
experiments.  The fact that Hermes has seen a robust (2--3\%) asymmetry
with a longitudinally polarized target suggests that they will see a
large asymmetry with a transversely polarized target (unless the
effect is entirely twist three -- e.g.,  $h_{L}\gg\delta q$).  This in
turn will lead to the first measurements of the nucleon's transversity
distribution and to new insight into the relativistic spin structure
of confined states of quarks and gluons.

\section{Conclusions}

A richer and more complex picture of the QCD bound states has emerged
since the 1987 renaissance precipitated by the EMC observation that
quarks carry only a small fraction of the nucleon spin.  We know much 
more about the nucleon's spin than we did back then.  We also know 
what to look for in the future:  we have a clear program for future 
measurement and analysis of the gluon helicity distribution, $\Delta 
g$, the quark transversity, $\delta q_{a}$, and the flavor 
decomposition of the quark spin ($\Delta \bar u$, $\Delta \bar d$, 
etc.)\ and a host of other related subjects, which I have not had time 
to discuss here.  This program involves several facilities and 
different energy regimes.  The polarized $ep$ collider we are 
considering at this workshop clearly has a central role to play.

\subsection*{Acknowledgments}
This work is supported in part by funds provided by the U.S.
Department of Energy (D.O.E.) under cooperative
research agreement \#DF-FC02-94ER40818.


\end{document}